\documentclass[twocolumn,aps,prl,showpacs]{revtex4}

\usepackage{graphicx}
\usepackage{graphics}
\usepackage{epsfig}
\begin{document}
\title{\bf\Large Effective restoration of the U$_A$(1) symmetry with temperature and density}
\author{\bf P. Costa}
\email{pcosta@teor.fis.uc.pt}
\author{\bf M. C. Ruivo}
\email{maria@teor.fis.uc.pt}
\author{\bf C. A. de Sousa}
\email{celia@teor.fis.uc.pt}
\affiliation{Departamento de F\'{\i}sica, Universidade de Coimbra,
P-3004-516 Coimbra, Portugal}
\author{\bf Yu. L. Kalinovsky\footnote{Permanent Address: Laboratory of Information Technologies, Joint Institute for Nuclear Research, Dubna, Russia}}
\email{kalinov@qcd.phys.ulg.ac.be}
\affiliation{Universit\'{e} de Li\`{e}ge, D\'{e}partment de Physique 
B5, Sart Tilman, B-4000, LIEGE 1, Belgium}
\date{\today}
\begin{abstract}
 
We investigate the full U(3)$\otimes$U(3) chiral symmetry restoration, at finite temperature and density, on the basis of a quark model which incorporates the most relevant properties of QCD in this context: explicit and spontaneous breaking of chiral symmetry and axial U$_A$(1) symmetry breaking. A specific lattice-inspired behavior of the topological susceptibility, combined  with the convergence of chiral partners, signals the onset of an effective chiral symmetry restoration. The results suggest that the axial part of the symmetry is restored before the possible restoration  of the full U(3)$\otimes$U(3) chiral symmetry  can occur. This conclusion is valid in the context of both finite temperature and density.

\end{abstract}
\pacs{11.10.Wx, 11.30.Rd, 14.40.Aq, 24.85.+p}

\maketitle


Restoration of symmetries and deconfinement are expected to occur  under extreme conditions (high density and/or temperature) that may be achieved in ultra relativistic heavy-ion collisions or in the interior of neutron stars.
An interesting open question  is whether  both  chiral SU$(N_f)\otimes$SU$(N_f)$  and axial U$_A$(1) symmetries are restored and which observables could carry information about the possible restorations.

The QCD Lagrangian is symmetric under SU$(N_f)\otimes$SU$(N_f)$ and U$_A$(1) transformations, which implies the occurrence of nine Goldstone bosons for $N_f=3$. It is well known that the U$_A$(1) symmetry does not exist at the quantum level being explicitly broken by the axial anomaly \cite{Weinberg}. 
The breaking of the U$_A$(1) symmetry can be  described at the semiclassical level by instantons \cite{t Hooft}, which gives a mass to $\eta^{\prime}$ in the chiral limit and, away from this limit, contributes to lift its mass to about $1$ GeV. 
The U$_A$(1) anomaly   is also responsible  for flavor mixing, that has the effect   of lifting the degeneracy between, for instance,  $\pi^0$ and $\eta$ which are degenerate in U(3), even in the presence of explicit symmetry breaking.  Due to the strong flavor mixing between $(q\bar q)_{\rm ns}\,=\frac{1}{\sqrt 2}(u \bar u\,+\,d \bar d)$ and $(q\bar q)_{\rm s}\,=\,s \bar s$, the ideal mixing is not achieved,  with the consequent violation of the Okubo-Zweig-Iizuka [OZI] rule for both  pseudoscalar and scalar mesons. So, the restoration of the  U$_A$(1) symmetry should have   observables effects both on the scalar and pseudoscalar meson spectrum and on the phenomenology of meson mixing angles as well. 
 
So far as  the restoration of symmetries is concerned, and assuming that both SU(3)$\otimes$SU(3)  and U$_A$(1) chiral symmetries are restored, two scenarios discussed by Shuryak are usually considered \cite{shuryak}: in scenario 1, $T_c<<T_{U(1)}$ and  the complete U(N$_f$)$\otimes$U(N$_f$) chiral symmetry is   restored well inside the quark-gluon plasma region; in scenario 2,  $T_c\approx T_{U(1)}$. The behavior of $\eta^{\prime}$ or of related observables in hot and dense medium \cite{kapusta}, like the topological susceptibility, might help to decide between these scenarios. The topological susceptibility is essential for the study of the  U$_A$(1) problem. In fact, this quantity in SU(3) gauge theory can be related to the finite $\eta'$ mass through the Witten-Veneziano formula \cite{veneziano} and several studies have been  done linking the behavior of the topological susceptibility with the restoration of the U$_A$(1) symmetry with the temperature.
Lattice results, both in pure color SU(3) gauge theory \cite{latticeChu,lattice} as well as for the unquenched case \cite{latticeChu2}, indicate a sharp drop of the topological susceptibility by an order of magnitude at the deconfining temperature, showing an apparent restoration of the U$_A$(1) symmetry. Preliminary lattice results give an  indication that there is a drop of the topological susceptibility with increasing baryonic matter \cite{bartolome}. 

A criterion to identify an effective restoration of  chiral symmetry  is to look for the degeneracy of the respective chiral partners. 
It is generally assumed that, whether SU(3)$\otimes$SU(3) and U$_A$(1) symmetries are restored at the same point or one precedes the other, when finally they get both restored $a_0$ and $\sigma$ mesons become degenerate with the $\pi^0$ and $\eta$ mesons.

This leads to a closer look on the scalar sector, $J^P=0^+$, which has been under intense investigation over the past few years \cite {Beveren}.
The main problem concerning these mesons is that there are too many light scalars below 1 GeV. One can accept that the two isoscalars $\sigma$ ($f_0(500)$) and $f_0(980)$  \cite {pdb} as well as the isovector $a_0(980)$ and the isospinor $K_0^*(800)$ \cite {E791} scalars   are enough candidates to fill up a nonet of light scalars. Although it is accepted that large 4-quarks and meson-meson components \cite {Close} are necessary to explain  this nonet, here  we shall assume  a $q \bar q$ structure for the scalar
mesons which are relevant to study the restoration of both chiral and axial symmetries. Recently, Dai and Wu \cite{Day} claimed that ($\sigma,\,f_0,\,a_0$ and $K^*_0$) can be chiral partners of the pseudoscalar nonet  ($\eta,\eta^\prime,\pi^0$ and $K$).


We perform our calculations in the framework of an extended  SU(3) Nambu--Jona-Lasinio model [NJL]. The Lagrangian density that includes the 't Hooft instanton induced interaction term that breaks the U$_A$(1) symmetry  \cite{RKS,kuni} is of the form:
\begin{eqnarray}
{\mathcal L\,}&=& \bar q\,(\,i\, {\gamma}^{\mu}\,\partial_\mu\,-\,\hat m)\,q\nonumber\\
&+& \frac{1}{2}\,g_S\,\,\sum_{a=0}^8\, [\,{(\,\bar q\,\lambda^a\, q\,)}
^2\,\,+\,\,{(\,\bar q \,i\,\gamma_5\,\lambda^a\, q\,)}^2\,] \nonumber\\
&+& g_D\,\{\mbox{det}\,[\bar q\,(1+\gamma_5)\,q] +\mbox{det}
\,[\bar q\,(1-\gamma_5)\,q]\}. \label{1}
\end{eqnarray}
Here $q = (u,d,s)$ is the quark field with three flavors, $N_f=3$, and
three colors, $N_c=3$, $\hat{m}=\mbox{diag}(m_u,m_d,m_s)$ is the current
quark mass matrix and $\lambda^a$ are the Gell--Mann matrices, a = $0,1,\ldots , 8$, ${ \lambda^0=\sqrt{\frac{2}{3}} \, {\bf I}}$.
The coupling constants have nontrivial dimensions: 
$g_S\,\propto [\rm {mass}]^{-2}$ and $g_D\,\propto [\rm {mass}]^{-5}$.
As usually, we introduce a three-momentum cutoff $\Lambda$  to regularize divergent integrals.
By using a standard bosonization procedure,  an effective meson action is obtained, leading to gap equations for the constituent quark masses and to meson propagators from which several observables are calculated.
The model parameters are given in Refs. \cite{RKS,costaI,costaB,costabig,costaD}.

In what follows we will concentrate first on the restoration of the symmetries at zero density and finite temperature. In the chiral models, when the coefficient of the anomaly term is constant, although there is a decrease of  the  U$_A$(1) violating quantities, this symmetry is not restored \cite{costabig,bielich2} due to the fact that the strange quark condensate does not decrease enough \cite{Bielich}. However, an effective restoration may be   achieved   by assuming that the coefficient of the anomaly term in the Lagrangian ($g_D$ in the present case) is a dropping function of the temperature.

Motivated by phenomenology arguments, a temperature dependence of the anomaly coefficient, in the form of a decreasing exponential, was firstly proposed by Kunihiro \cite{kuni}, in the framework of the present model. 
A different approach to this problem appears in more recent works, that make use of lattice results \cite{lattice} for the topological susceptibility as input \cite{Bielich,Ohta}. 

In the present work we consider the anomaly  coefficient $g_D$ as a dropping function of temperature, following the  methodology of Ref. \cite{Ohta}, and extract its temperature dependence from the lattice results for the topological susceptibility  \cite{lattice}.
As we can see in Fig. 1, upper panel, the topological susceptibility, $\chi$, is a Fermi function, which is almost constant until $T\approx150$ MeV, the lattice result for the critical temperature \cite{lattice1, lattice2}. 
The discussion of possible restoration of axial and chiral symmetries is based on the analysis of the behavior of chiral partners, as in Ref. \cite{Bielich}, being complemented by the study of the mixing angles.
%
\begin{figure}[t]
\vspace{-0.6cm}   
  \begin{center}
    \includegraphics[width=0.46\textwidth]{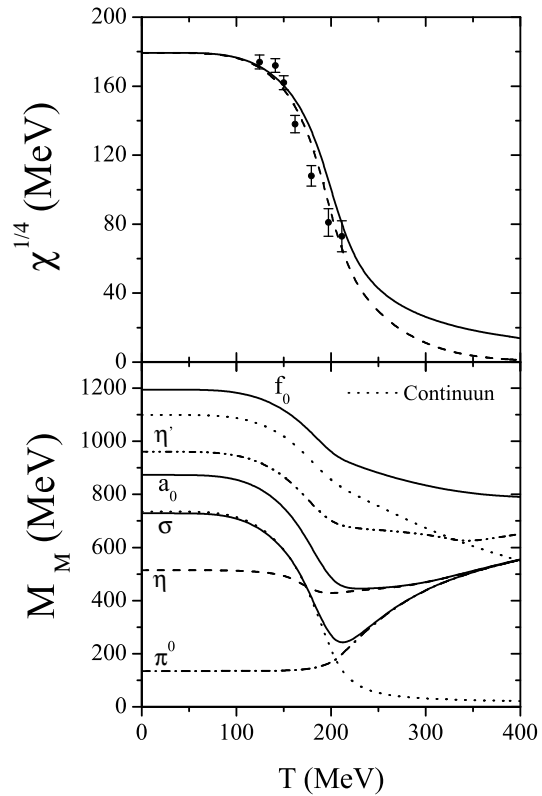}
  \end{center}
\vspace{-0.75cm}   
\caption{Topological susceptibility (upper panel) from lattice data plotted with error bars \cite{lattice}. The solid (dashed) line represents our fitting with constant (temperature dependent) $g_D$. Temperature dependence of the meson masses (lower panel), as well as that of the continuum thresholds $2M_u$ and $2M_s$.}
\label{fig:temp}
\end{figure}

Our results, concerning the restoration of the chiral phase transition, at zero density and finite temperature, indicate, as usual, a smooth crossover. Contrarily to what is reported in Refs. \cite{kuni,Bielich} the temperature dependence for $g_D$ used here and in Ref. \cite{Ohta}, does not strengthen the chiral phase transition, the results for the critical temperature are around the same value as for a constant coefficient. At temperatures around $T\approx200$ MeV the mass of the light quarks drops to the current quark mass, indicating a washed-out crossover from the chirally broken to approximate chirally symmetric phase. The strange quark mass presents also a significant decrease  in this temperature range, however even at $T = 400$ MeV it is still 2 times the strange current quark mass. So we can say that the chiral symmetry shows a slow tendency to get restored in the $s$ sector.

In fact, as $m_u=m_d<m_s$, the (sub)group SU(2)$\otimes$SU(2) is a much better symmetry of the Lagrangian (1). 
So, the effective restoration of the above symmetry  implies the degeneracy between the  chiral partners $(\pi^0,\sigma)$ and $(a_0,\eta)$ around $T\approx250$ MeV (see Fig. 1, lower panel).
For temperatures  at $T\approx350$ MeV both $a_0$ and $\sigma$ mesons become degenerate with the $\pi^0$ and $\eta$ mesons, showing, as explained below, an effective restoration of both chiral and axial symmetries.
In fact, without the restoration of U$_A$(1) symmetry, the $a_0$ mass was moved upwards and never met the $\pi^0$ mass, the same argument being valid for the $\sigma$ and $\eta$ mesons.
We remember that the determinantal term acts in an opposite way for the scalar and pseudoscalar mesons.
So, only after the effective restoration of U$_A$(1) symmetry we can recover the SU(3) chiral partners $(\pi^0,a_0)$ and $(\eta,\sigma)$ which are now all degenerated.
However, the $\eta^\prime$ and $f_0$ masses do not yet show a clear tendency to converge in the region of temperatures studied, this absence of  convergence being probably due to the fact that, in the region of temperatures above $T\approx350$ MeV, those mesons are purely strange and the chiral symmetry in the strange sector is far from being restored.

A complementary analysis of the temperature dependence of the mixing angles, allowing for a better understanding of the meson behavior through the evolution of the meson quark content, provides further indication of the restoration of the axial symmetry: $\theta_S$ starts at $16^{\circ}$ and goes, smoothly, to the ideal mixing angle $35.264^{\circ}$ and $\theta_P$ starts at $-5.8^{\circ}$ and goes to the ideal mixing angle $-54.7^{\circ}$. This means that flavor mixing no more exists.
In fact, (see Fig. 1, lower panel) we found that the $a_0$ meson is always a purely non strange quark system while the $\eta$ meson, at $T = 0$ MeV, has a strange component and becomes purely non strange when $\theta_P$ goes to $-54.7^{\circ}$ at $T \approx 250 $ MeV. At this temperature they start to be degenerated. Concerning the SU(2) chiral partner ($\pi^0,\sigma$), at $T=0$ MeV, $\pi^0$ is always a light quark system and the $\sigma$ meson has a strange component,  but becomes purely non strange when $\theta_S$ goes to $35.264^{\circ}$ at $T \approx 250 $ MeV. Summarizing,  we  conclude that at $T\approx350$ MeV the $\pi^0$, $\sigma$, $\eta$ and $a_0$ mesons become degenerated, the OZI rule is restored and  $\chi$ goes asymptotically to zero (Fig. 1, upper panel). These results indicate an effective restoration of the U$_A$(1) symmetry.

It should be noticed that our analysis of the restoration of symmetries is based on the degeneracy of chiral partners that occurs in a region of temperatures where the mesons are no more bound states (they dissociate in $q \bar q $ pairs at their respective Mott temperatures \cite{RKS,costabig}). Moreover, the mesons $\eta'$ and $f_0$ are $q \bar q $ resonances from the beginning and its description is unsatisfactory.  

Due to recent studies on lattice QCD at finite chemical potential \cite{bartolome} it is tempting to investigate also the restoration of the U$_A$(1) symmetry at finite  density and zero temperature. 
The disadvantage in this case is that there are not yet firmly established  lattice results for the density dependence of the topological susceptibility, to be used as input to model this  quantity, so we have to extrapolate from our previous results and proceed by analogy. We postulate a dependence for the topological susceptibility formally similar to the temperature case.  Here we present an example  (see Fig. 2, upper panel).

\begin{figure}[t]
\vspace{-0.6cm}   
   \begin{center}
       \includegraphics[width=0.52\textwidth]{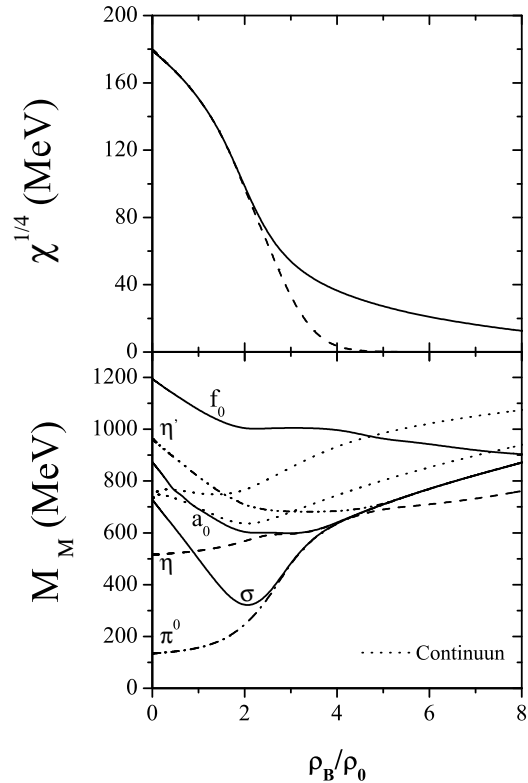}
   \end{center}
\vspace{-0.75cm}   
\caption{Topological susceptibility (upper panel): the solid (dashed) line represents our fitting with constant (density dependent) $g_D$. Meson masses (lower panel) as functions of density. The dotted lines indicate the density dependence of the limits of the Dirac sea continua, defining $q\bar q$ thresholds for $a_0$ and $\eta^\prime$ mesons. }
\label{fig:dens}
\end{figure}

To study the behavior with the density, we  will consider quark matter simulating "neutron" matter. This "neutron" matter is in chemical equilibrium, maintained by weak interactions and with charge neutrality, and undergoes a first  order phase transition (for details see Refs. \cite{Buballa,costaI,costabig}).

The behavior of non strange and strange quark masses, although presenting details different from the finite temperature case, have qualitative similarities that lead to the same conclusion concerning restoration of chiral symmetry: it is effectively restored in the non strange sector, but the same does not happen in the strange sector. Also similarly to the finite temperature case, the chiral transition is not affected by the density dependence of the anomaly coefficient, but, concerning the density effects on  the meson mass spectrum and on mixing angles, although up to a certain value of the density the results are  qualitatively similar to the finite temperature case, some new aspects, that will be discussed in the sequel, appear at high densities.

Analyzing the mixing angles we observe that the behavior of $\theta_S$ is similar to the non zero temperature case: it starts at $16^{\circ}$ and increases up to the ideal mixing angle $35.264^{\circ}$. A different behavior is found for the angle  $\theta_P$, that changes sign at $\rho_B\approx4\rho_0$: it  starts at $-5.8^{\circ}$ and goes to the ideal mixing angle $35.264^{\circ}$, which, as it will be seen, leads to a change of identity between $\eta$ and $\eta'$. We think this result might be a useful contribution for the understanding of the somewhat controversial question: at high densities or temperature will the pion degenerate with $\eta$ or $\eta'$? The answer is related to the possible change of sign of the pseudoscalar mixing angle. We found that the reason for the change of sign, as compared to the previous case, is related to a different decrease of the strange quark mass and depends on the amount of strange quarks in the medium \cite{costanew}.
 
The meson masses, as function of the density, are plotted in Fig. 2, lower panel. 
The SU(2) chiral partners ($\pi^0,\sigma$) are now always bound states. The pion is a light quark system for all range of density and the $\sigma$
meson has a strange component at $\rho_B = 0$ but becomes purely non strange when $\theta_S$ goes to $35.264^{\circ}$, at $\rho_B=3\rho_0$.
At this density the mesons become degenerated. This behavior is similar to the non zero temperature case.

The surprise happens with the SU(2) chiral partners ($\eta,a_0$). The $a_0$ meson is always a purely non strange quark system, as in the temperature case. For $\rho_B<0.8\rho_0$ $a_0$ is above the continuum and, when $\rho_B\geq0.8\rho_0$, $a_0$ becomes a bound state. However, the $\eta$ meson has an interesting different behavior. At $\rho_B = 0$, the $\eta$ has a strange component and, as the density increases, $\eta$ becomes degenerated with $a_0$ at $4.0\rho_0\leq\rho_B\leq4.8\rho_0$ as expected. In this range of densities, ($\eta,a_0$) and ($\pi^0,\sigma$) are all degenerated. Suddenly the $\eta$ mass separates from the others becoming a purely strange state. This is due to the behavior of the $\theta_P$ that changes the sign and goes to $35.264^{\circ}$ at $\rho_B\approx4.9\rho_0$. On the other hand, the $\eta'$, that starts as an unbounded state and becomes bounded at $\rho_B>3.0\rho_0$, turns into a purely light quark system and degenerates with $\pi^0$, $\sigma$ and $a_0$ mesons. So, the $\eta$ and the $\eta'$ change identities.  Consequently, contrarily to results with temperature, $\pi^0$ and $\eta^\prime$ are now degenerate. The topological susceptibility tends to zero asymptotically. We notice that, at high densities, and  differently from the non zero temperature case,  all the mesons are bound states. 
Taking into account the presented arguments, we conclude that the U$_A$(1) symmetry is effectively restored at $\rho_B>4\rho_0$.

In summary, the present investigation explores a framework to study effective chiral and axial symmetry restoration  with temperature and density. 
We implement  a criterion which combines a lattice-inspired behavior of the topological susceptibility with the convergence of appropriate chiral partners. 
We observe that our results are compatible  with scenario 1 of Shuryak \cite{shuryak}.
In fact, the signals for the restoration of the U$_A$(1) symmetry occur at a moderate temperature (density), where signals of the full restoration of  U(3)$\otimes$U(3) are not yet visible. 
However, the role of U$_A$(1) symmetry for finite temperature, and mainly for finite density media, has not been so far investigated and this question is still controversial and not settled yet.
We hope that new studies, especially lattice based and experimental ones, can finally clarify it. 

We thank Bartolome All\'{e}s for information about the present stage of lattice calculations of the topological susceptibility  in baryonic matter.

Work supported by Grant No. SFRH/BD/3296/2000 (P.C.), by Grant No. RFBR 03-01-00657, Centro de F\'{\i}sica Te\'orica and GTAE (Yu.L.K.), and by FEDER/FCT under projects POCTI/FIS/451/94 and POCTI/FNU/50326/2003. 


\end{document}